\documentclass[twocolumn]{revtex4}
\usepackage{graphicx}
\usepackage{dcolumn}
\usepackage{bm}
\usepackage{amsfonts}

\begin{document}
\title[Simulation of dark lanes in post--flare supra--arcade]{Numerical simulation of dark lanes in post--flare supra--arcade}

\author{Andrea Costa, acosta@mail.oac.uncor.edu}
\affiliation{Instituto de Astronom\'\i a Te\'orica y Experimental (IATE, C\'ordoba, Argentina) \\
Consejo Nacional de Investigaciones Cient\'\i ficas y T\'ecnicas (CONICET) \\ Facultad de Ciencias Exactas, F\'\i sicas  y Naturales (UNC-Universidad Nacional de C\'ordoba, Argentina)}
\author{Sergio Elaskar, selaskar@yahoo.com}
\affiliation{Facultad de Ciencias Exactas, F\'\i sicas  y Naturales (UNC, Argentina) \\
Consejo Nacional de Investigaciones Cient\'\i ficas y T\'ecnicas (CONICET) }
\author{Carlos A. Fern\'andez,  ultritas@yahoo.com.ar}
\affiliation{Facultad de Ciencias Exactas, F\'\i sicas  y Naturales (UNC, Argentina)}
\author{Guadalupe Mart\'\i nez, gmartinez@iafe.uba.ar}
\affiliation{Instituto de Astronom\'\i a y F\'\i sica del Espacio (IAFE, Buenos Aires, Argentina)}

\begin{abstract}
We integrate the MHD ideal equations to simulate dark void sunwardly moving structures in post--flare supra--arcades. We study the onset and evolution of the internal plasma  instability to compare with observations  and to gain insight into physical processes and characteristic parameters of these phenomena. The numerical approach uses a finite-volume Harten-Yee TVD scheme to integrate the 1D$\frac{1}{2}$ MHD equations specially designed to capture supersonic flow discontinuities.  The integration is performed in both directions, the sunward radial one and the transverse to the magnetic field. For the first time, we  numerically reproduce  observational  dark voids described   in Verwichte et al. (2005). We show that the dark tracks are   plasma vacuums   generated by the bouncing and interfering of shocks and expansion waves, upstream an initial slow magnetoacoustic   shock  produced by a localized deposition of energy modeled with a pressure perturbation. The same pressure perturbation produces  a transverse to the field or perpendicular magnetic  shock  giving rise to nonlinear waves that compose the kink--like  plasma void structures,  with the same functional sunward decreasing phase speed  and constancy  with height of the  period,   as those  determined by the observations. 

\end{abstract}
\maketitle

\section{Introduction}
Dark sunward sinuous lanes moving  along a fan of rays  above post--flare loops towards a supra--arcade   with a decelerating speed  in the range of $\sim (50 - 1000) \ $km s$^{-1}$  were firstly described by McKenzie and Hudson  \cite{mck}, using the Yohkoh Soft X--ray Telescope. Further observations showed that 
the fraction of solar flare long--duration event  arcades possessing fans is not small, and that  downflows above such arcades are not uncommon (McKenzie \citealp{mck2}). 
The phenomena 
were extensively studied by several authors (Wang et al. \citealp{wan}; Innes et al. \citealp{inn}a,b;  Svestka et al. \citealp{sve}; Asai et al. \citealp{asa}; McKenzie and Savage \citealp{mck3}) and different theoretical explanations have been offered to give account of the physics of the underlying process.  McKenzie   \cite{mck2} dismissed the interpretation that X--ray voids could be explained as evidence of blobs of cool material separated from the main body of a CME  falling gravitationally  back to the surface of the Sun. This author supported the explanation that flux tubes linking an above current sheet retract into the arcade under the force of magnetic tension (see also Wang et al. \citealp{wan}). The down moving structures  observed at $40-60 \ $Mm heights above the top of the arcade indicated that the reconnection site must be located above the flare arcade. 
Sheeley and Wang \cite{she} interpreted the phenomenon as the top of collapsing loops with void tails in the wakes where reconnection occurs. Asai et al. \cite{asa} proposed, based on RHESSI observations, that dark lanes are sunward voided flows generated by the reconnection process developed by a 
 current sheet at  the supra--arcade. 

Verwichte et al. (\citealp{ver}, hereafter VNC) analyzed  transverse to the magnetic field oscillations associated with sunward dark lanes in the post--flare supra--arcade of NOAA active region 9906 on the $21$st of April $2002.$ They interpreted   two oscillating in phase edges  (C and D in VNC)  as fast surface magnetoacoustic kink wave trains guided by the ray  structure. 
They founded that the phase speeds and the displacement amplitudes decrease as they propagate downwards and derived a function to fit the amplitude of the oscillating perturbation at a given height. The period remained constant with height. As a function of time,  the phase speeds decreased and periods increased. 

In this Letter, based on the following arguments, we  propose a new scenario to numerically give account of the phenomenology described in  VNC: 1) the observational  phase speed magnitude in VNC is several hundreds of km s$^{-1}$ lower than the Alfv\'en  estimations, this could suggest that slow speed sunward processes, rather than fast ones, are at work; also, 2) an impulsive energy deposition or pressure pulse can lead to lower speed void features  obtained as a result of nonlinear interacting slow shocks, i.e.,  where the magnetic field plays the role of being the wave guide of a fundamentally hydrodynamic shock  (Fern\'andez et al. \citealp{fer}). 
This  fact is a starting supposition  to  simulate  the dynamics along  the magnetic field lines.  3) The sunward dynamics is differentiated,  from the transverse one because conductive energy transport along  field lines  is far more effective than across them, thus, we assume independence between the dynamics of  two main directions: the sunward and the transverse to the magnetic field one. 4) As the impulsive triggering mechanism acts in all directions, a perpendicular magnetic shock can develop  in the transverse direction.  
 In conclusion, our modeling is accomplished supposing that a pressure pulse  applied  to the coronal inhomogeneous medium, structured by a main magnetic field,  is a triggering mechanism of two fundamental independent shock schemes, an hydrodynamic one and a magnetic perpendicular one. On each direction the interaction of the nonlinear waves will give rise to the different patterns we are interested in. 
We emphasize that if the simulations  reproduce  the observational data, e.g., the kink--like features, the temporal cadences, the phase speed and sunward pattern functional decreases, in the frame of a few reasonable suppositions, this implies that the overall  scheme could be at the bases of the underlying  physical processes. Even when a full numerical simulation is required,  this  two 1D $\frac{1}{2}$ modeling, is a useful first approach, certainly, if due to its simplification  and to the accuracy in the observational description suggests that it  had captured  important physical  processes.

\section{Numerical method}

To  simulate  time dependent MHD flows we use a method developed by  us (Elaskar et al. \citealp{ela}). The equations used are the continuity, momentum, energy and magnetic induction equations together with the state equation forming   a system of partial hyperbolic  differential equations. The numerical approach consists of an approximate Riemann solver  with the TVD scheme proposed by Yee for gasdynamics flows (Yee et al.  \citealp{yee}). The ``eigensystem'' technique presented by Powell \cite{pow} was also  used with the eigenvector normalization proposed by Zarachay et al. \cite{zar}. The  accuracy of the technique was verified by means of simulations of the Riemann problem introduced by Brio and Wu and the Hartmann flows (Maglione et al. \citealp{mag}).

\section{Results and discusion}

In VNC,  to study supra--arcade ray structures the authors chose a rectangular subfield of parallel $(x)$ and perpendicular  $(y)$ to the supra-arcade ray  coordinates where the structures were visible. They could describe the oscillatory motions of four dark  transverse displacements  for a fixed vertical coordinate  as a function of time, i.e. $\xi_{y}(t); x=x_{o}=45.6$Mm. 
 Also, they displayed  the intensity variations  of the ray parallel coordinate as a function of time i.e., $I_{x}(t)$ averaged over  a segment of the $y$ coordinate,  $\Delta y \sim 4$Mm,  (Fig.~1-2 in VNC).
They found that the period $\tau$  remains constant with height but the amplitude and phase speed decrease as the wave travels sunwards. Assuming a transverse perturbation of the form $
\delta \xi_{y}(x,t)=A \cos[\frac{2 \pi t}{\tau}-\phi] , $
  $t$ is the time, and $x$ the height, the functions that fit the amplitude and  the  phase speed are respectively  
\begin{equation}
A(x)=A_{o}\exp[\frac{x-x_{o}}{L_{A}} ]  \  \ \ \ V_{ph}=V_{pho}\left( 1-\frac{(x-x_{o})}{L_{v}}\right) ^{-1} \label{Amp}
\end{equation}
with  correspondent  parameters given in Table 1.

To  reproduce the observations  we schematically chose two set of  characteristic values in VNC: edges C and D, which  seem to have  different structure. Edge C's tail has a kink--like structure and edge D, more irregular, resembles a sausage--like structure in  part of its tail.   Table 1 shows the observational  wave parameters  in VNC  used to adjust the numerical simulation.

\begin{figure}
\begin{center}
 \includegraphics[width=7.5cm]{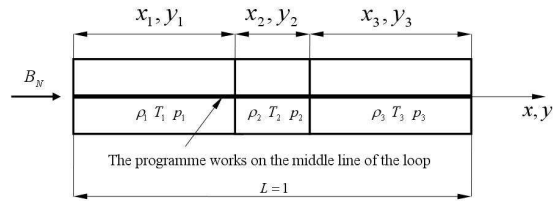}
 \end{center}
\caption{Scheme of the loop description. }
  \label{fig:cero}
   \end{figure}

To perform the numerical simulation the physical  quantities are expressed in a dimensionless form. As in VNC the coordinate $x$ represents the sunward direction and the $y$ coordinate the transverse to the magnetic field one, as seen by the line of sight.
 Figure~\ref{fig:cero} shows a scheme of the numerical domain  of  the  two  1D$\frac{1}{2}$  independent integrations performed, i.e.  two main initial shock  wave fronts triggered by the same  pressure perturbation. One directed sunwards,  parallel to the field lines, and the other  directed  transversally to them.  The composition of the two processes will give the  overall description. The details of the energy deposition mechanism are not taken into account here e.g., reconnection processes as the one suggested by Asai et al. \cite{asa}.

\subsection{Transverse perpendicular magnetic shocks}

Traditionally it is expected that the observations are due to the direct action of boundary effects between the void and the surrounding medium. Our simulations show that the observational void patterns can also be obtained as a result of  nonlinear evolution of waves triggered by an instantaneous  pressure pulse. It is required, however, that the media has density inhomogeneities  aligned with the magnetic field, in order to produce the interferences upon reflection that are responsible of sustaining the void lanes. The initial perturbation, seen as a transverse to the field line phenomenon, produces opposite magnetic shock waves that bounce in the lateral and denser boundary medium composing  an oscillatory pattern of nonlinear wave interactions whose evolution leads to  an internal defined  central vacuum region. 
The effect of the density inhomogeneities -which  produce partial reflections and absorptions of waves, e.g, it was estimated  that  density values can rise at least between two and five times from  the void to  the denser medium (McKenzie and Hudson  \citealp{mck}; Innes et al. \citealp{inn})- are taken into account as  density steps   modeled by  reflecting boundaries at the extremes of the integration domain.   
 We performed several simulation  with different values of the non--dimensional boundary density, allowing partial or total rebounds of the waves, and we found that  the result  is robust in the sense that the oscillating pattern is conserved   for a density range going from 
 four to ten times the non--perturbed initial value. A  study of  sensitive to the initial and boundary conditions will be  matter of a new work together with a 1D and 2D simulation.
 A resulting simulation that accurately fits the edge C description in VNC  is shown in Fig.~\ref{fig:uno}a. The non--dimensional boundary density is assumed constant and equal to $5.$ The  color palette indicate differences in the density values, a partial   rebound of half of the momentum occurs at $y_{n}=0$ and $y_{n}=1$. 
 The darker  region corresponds to the resulting vacuum zone pattern, with vanishing values of the density, i.e. the non--dimensional vacuum density is $0.3$ while  the external medium value results $\sim 1.4$.   Note that the simulation assumes an initial  pressure pulse that triggers the whole pattern, thus the following process of wave interactions produces and sustains the  void cavity for a  time interval which  is in accordance with the observations, as it can be seen from the figure. 
 
 If the $y_{i}$ partition is modified, a sausage--like description can be  obtained as it is shown in  Fig.~\ref{fig:uno}b. This figure was adjusted with the sausage--like edge D data. The differences between Fig.~\ref{fig:uno}a and Fig.~\ref{fig:uno}b are due to the fact that in the first case  an asymmetric rebound occurs while in the second one the rebound is symmetric.
Kink--like structures seem more probable than the sausage--like ones because  the symmetric condition requirement  of the external inhomogeneous media is more improbable.
Figure~\ref{fig:uno} shows that the vacuum is immediately originated by the pulse and the interaction of the non--linear waves  composes and sustains a kink--like or a sausage--like pattern for a large period of time. 

The procedure of convertion into the dimensionless quantities and the relations that couple   the transverse and radial parameters are described in next subsection.

\subsection{Calibration}
\begin{figure*}
   \includegraphics[width=8.5cm]{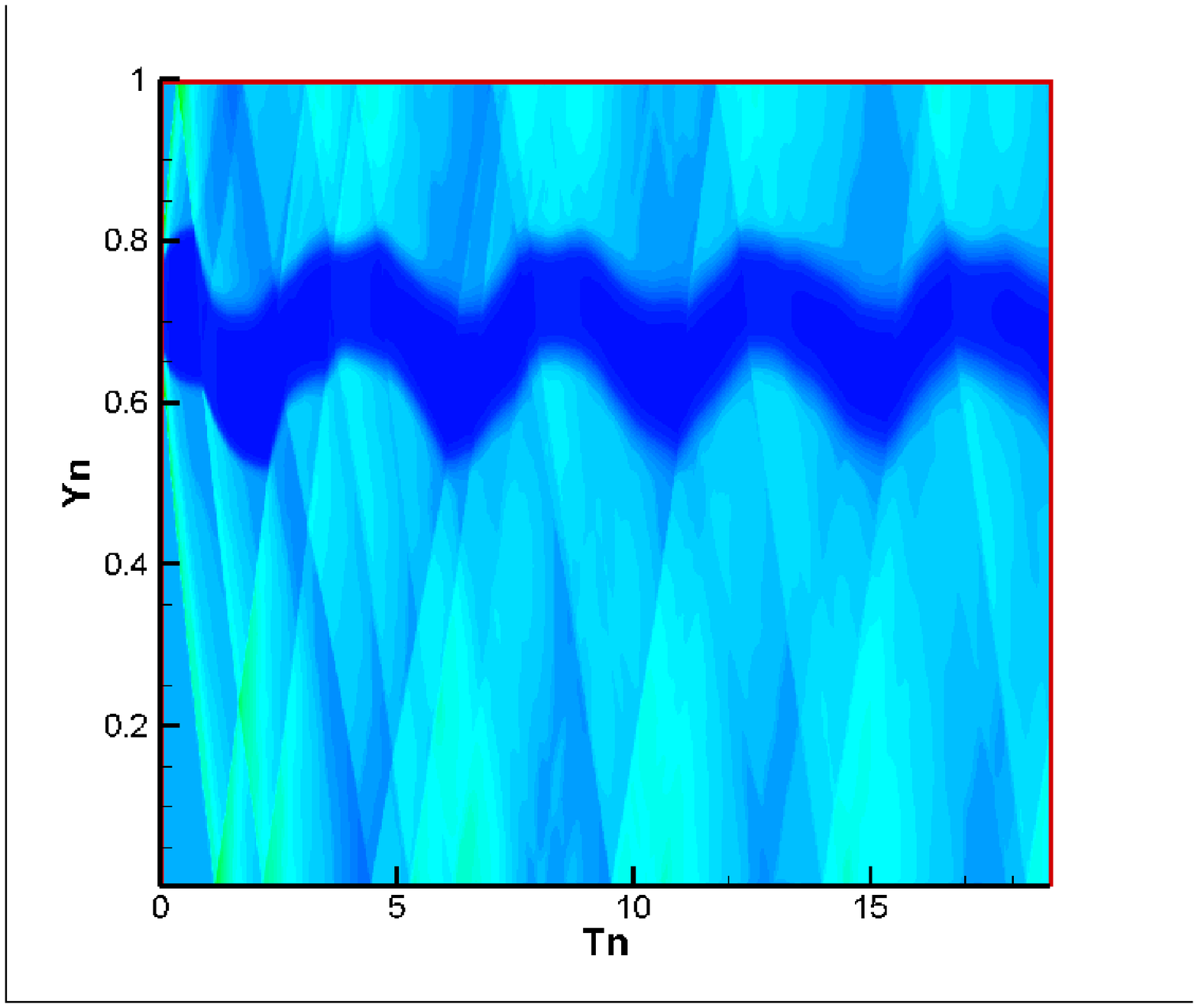}
  \hspace{0.5cm}
  \includegraphics[width=8.5cm]{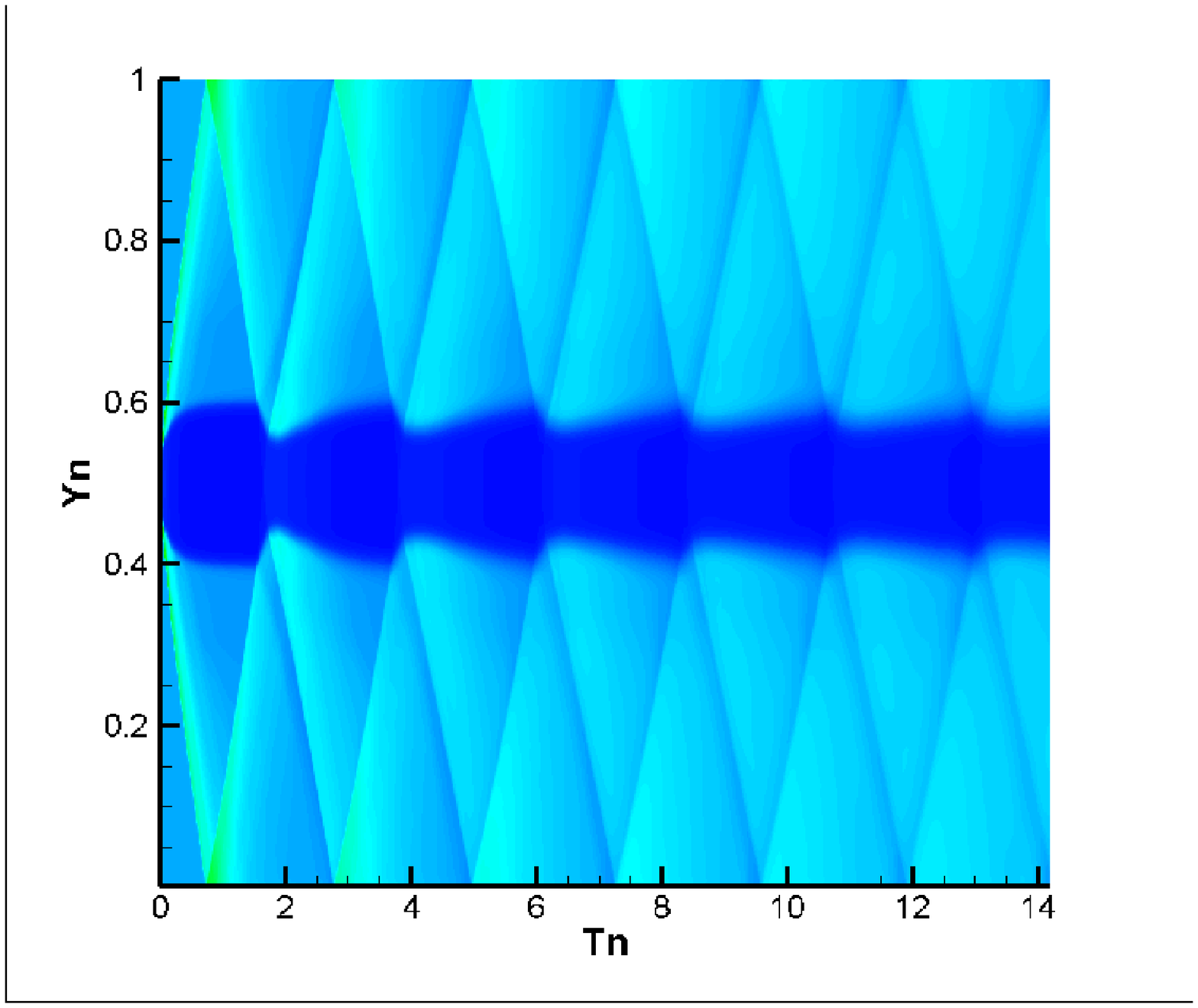}
\caption{a) Non--dimensional kink--like structure as edge C  in VNC. Asymmetric rebound of the lateral shock in the denser media. Dimensional values  obtained multiplying by $6040 \ $km and $29.7 \ $s.  b) Non--dimensional sausage--like structure as  edge D  in VNC. Symmetric rebound of the lateral shock in the denser media. Dimensional values  obtained multiplying by $2460 \ $km and $79.5 \ $s.}
  \label{fig:uno}
   \end{figure*}

The non--dimensional quantities are the ratios of a dimensional quantity and a reference one. 
We assume a reference value of the magnetic field and the gas pressure of $B_{ref}=10 \ $G and $P_{ref}=1.38 \ $Pa respectively. To derive the dimensional period, amplitude, density  and speed we proceeded as follows.
I) Using the first two lines of  non--dimensional values in Table 2, we obtained Fig.~\ref{fig:uno}.
II) 
From the figures  we determined the mean non--dimensional speed chosen as: 
\begin{equation}
  V_{n}=\frac{2A_{n}}{\tau_{n}} 
  \label{Vn}
\end{equation}
where $\tau_{n}$ and $A_{n}$ are the non--dimensional  periods and amplitudes respectively. The correspondent dimensional speed is:
\begin{equation}
  V = V_{n}V_{ref}=\frac{2A}{\tau} \  \longrightarrow  \ V_{ref}=\frac{2A}{\tau \ V_{n}}
  \label{V}
\end{equation}
where  the dimensional values
$\tau$ and $A$ are  taken from Table 1 as reported in VNC for  C and D edges. 
From  eq.~(\ref{Vn}) and eq.~(\ref{V}), taking into account  that $V_{ref}=B_{ref}/\sqrt{\mu_{o}\rho_{ref}}$, we obtain  the reference value for the density $\rho_{ref}=1.67 \ 10^{-11} \ $Kg$m^{-3}.$ 
Note that as $\tau = \tau_{n}\tau_{ref}$  and that from Fig.~\ref{fig:uno}a, $ \tau_{n}=4.5$ (or equivalently $ \tau_{n}=2.2$ from  Fig.~\ref{fig:uno}b) using the  edge C value $\tau=134  \ $s, taken  from Table 1 (or equivalently $ \tau=175 \ $s from  edge D) it results: $\tau_{ref}=29.7 \ $s for the kink structure and  $\tau_{ref}=79.5 \ $s for the sausage one. Using the same argument for the amplitude,  $A=A_{o} = A_{n}A_{ref}$, with $A_{o}$ taken from Table 1 and $A_{n}$ from Fig.~\ref{fig:uno}  we obtain $A_{ref}=6040 \ $km, for the kink structure, and $A_{ref}=2460 \ $km for the sausage one. These quantities are also used as the reference values  for the simulation in the sunward direction, together with the initial conditions shown in the last line of Table 2.

\subsection{Radial hydrodynamic shocks}


\begin{figure}
\begin{center}
  \includegraphics[width=8.1cm]{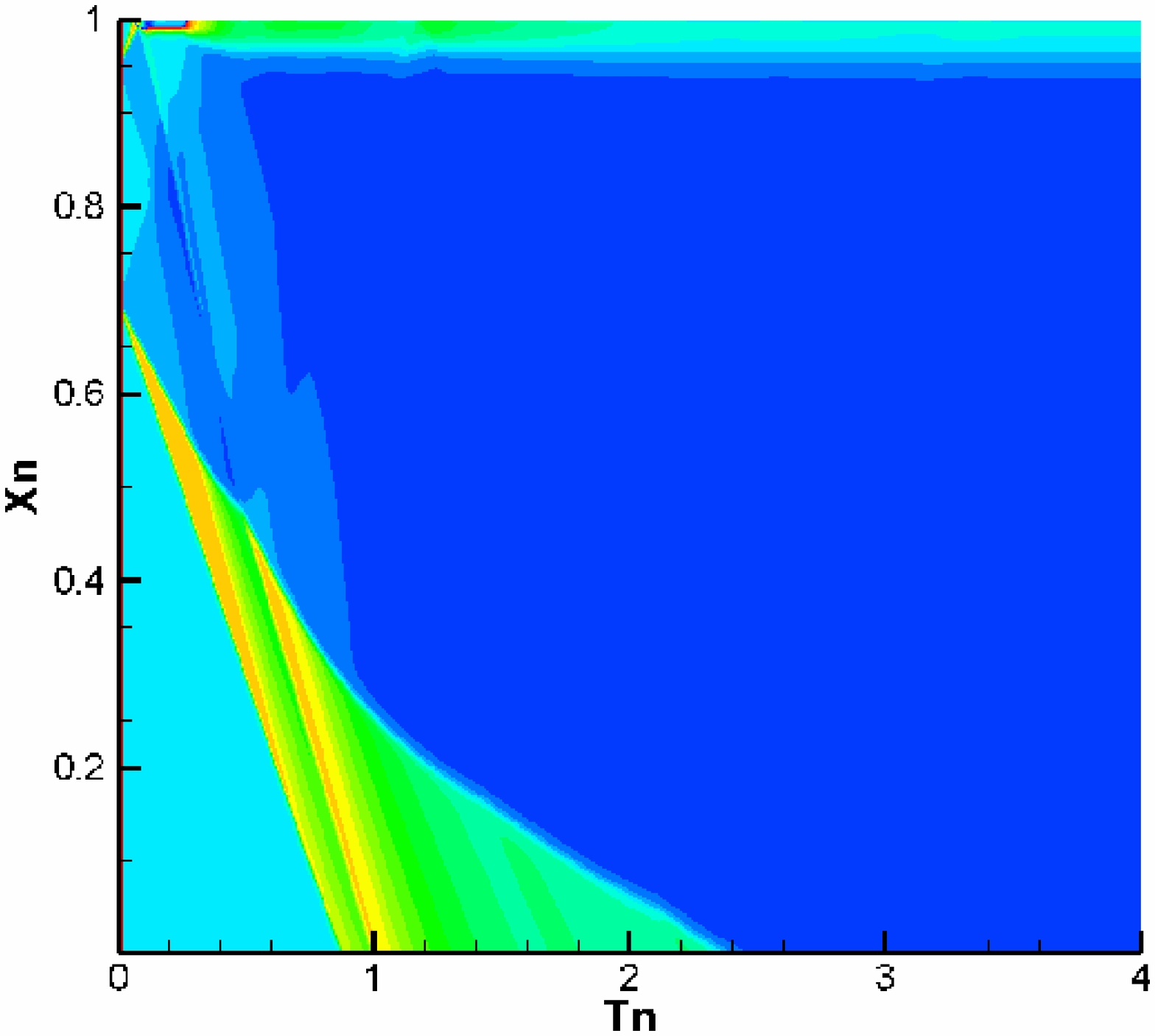}
               \includegraphics[width=7.5cm]{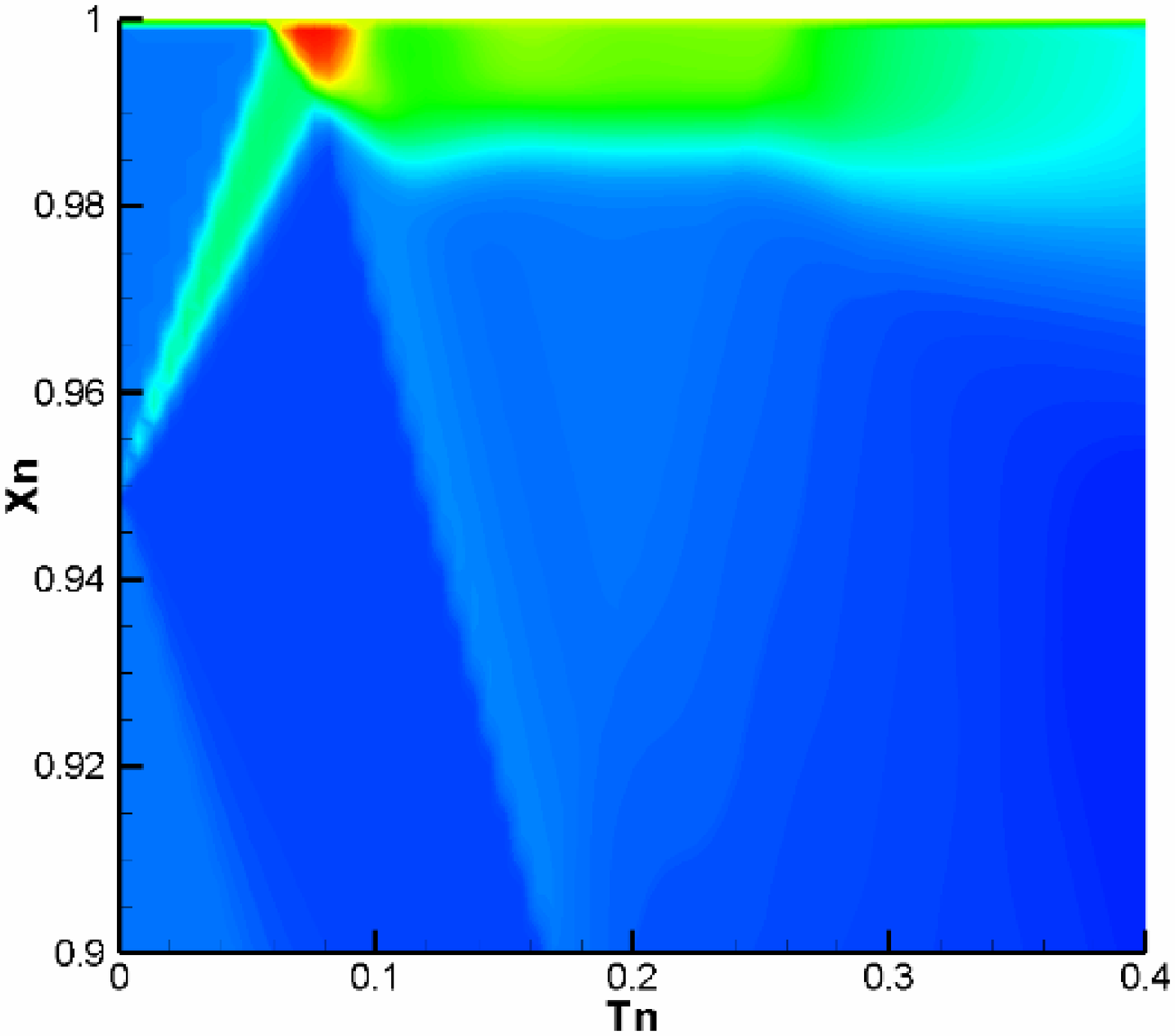}
              \includegraphics[width=7.5cm]{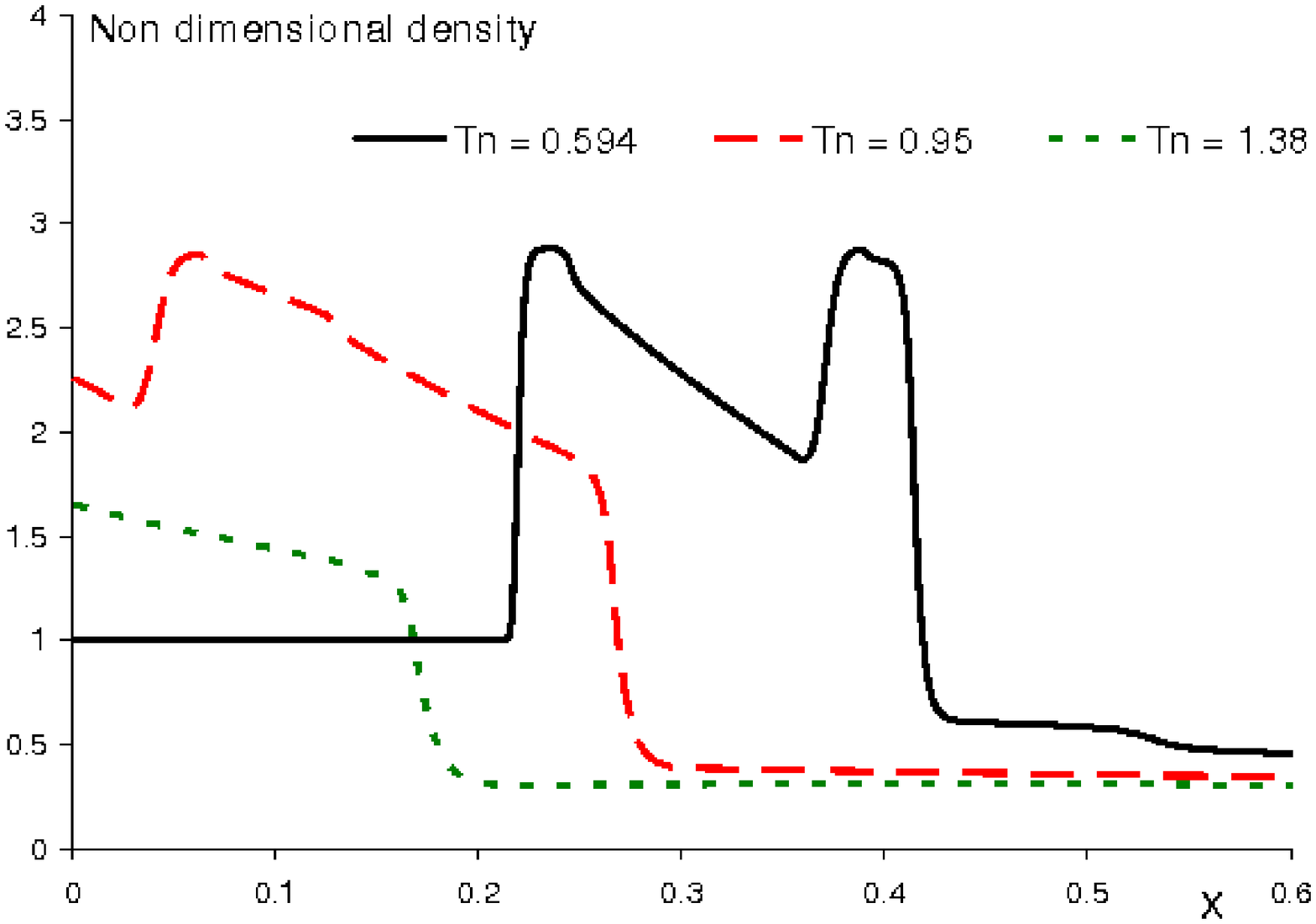}     
  \end{center}
\caption{(a)   Non--dimensional sunward displacement as a function of $x_{n}$ and $t_{n}$ for edge C. Dimensional values  obtained multiplying by $48.6 \ $Mm and $29.7 \ $s. Equivalent to Fig.~(2C) in VNC. (b) Zoom of (a) showing details of the upper rebound. (c) Density profile as a function of height for the times  $t=0.594, \ 0.95, \ 1.38.$ }
  \label{fig:dos}
   \end{figure}

In the radial direction  the same pressure pulse generates shocks traveling sunwards and outwards, along the magnetic field lines. The nonlinear interaction of outward rebound  waves and sunward absorbed ones  compose an overall descending void structure.  The pressure pulse is supposed to be associated with  a reconnection event implying that field lines should retract away from the reconnection site under the force of magnetic tension (McKenzie \citealp{mck2}). Thus, we  suppose that a partial rebound occurs upwardly near the reconnection site. Downwardly,  the background initial density is assumed constant and equal to the reference value. The sunward perturbation is absorbed by the non--perturbed media as it travels, however, a total absorption constraint is assumed at the sunward boundary. Our origin of coordinates in the radial direction is located at    
$x_{o}=45.6$Mm. In VNC $x_{o}$  is the radial coordinate indicating the fixed position from where the
motion is described, and Table 1 values were measured. We  chose $x_{t}=50$Mm  as the position of the triggering pressure pulse.
Thus, $x_{t}$ is also the  distance traveled by the void features, i.e.,  taking into account different  trajectories of the void heads in  VNC observations   this distance  results a good  assumption. The non--dimensional value of $x_{t}$ is $x_{on}=0.7$ and the non--dimensional initial density, $\rho_{n}=1,$ corresponds to the dimensional $\rho_{ref}$ value. 
 Figure~\ref{fig:dos}a is equivalent to Fig.~(2C) in VNC. 
 Note that the radial perturbation initially generates two shocks, one moving sunwards and the other upwards.  Figure~\ref{fig:dos}b is a zoom of  Figure~\ref{fig:dos}a showing the upward moving features, which  are forced to partially rebound at $x_{n}=1.$  The upwardly moving features in Fig.~\ref{fig:dos}b correspond to a shock front, recognized because of the density enhancement at the shock front, and a contact discontinuity wave, recognized because the pressure and the speed of the flow are not changed while the wave passes. Also, an expansion wave travels downwards, it is recognized because the density is diminished at its pass.  Figure~\ref{fig:dos}a shows that the interaction of  subsequent waves   composes a downward moving void as it is indicated in  by the color contrast, i.e. darker features correspond to lower values of the density. 
The clearest curve  that originates at $x_{n}=x_{on}=0.7,$  $t_{n}=0,$ and ends at  $x_{n}=0, $  $t_{n}=0.9,$ is the main downward shock, i.e. while crossing the shock front the density varies from  $\rho_{n}=1$ to $\rho_{n}=3.$ Later interactions compose a region of smooth expansion   that moves sunwards until it is absorbed at $x_{n}=0, $  $  t_{n } \sim \ (0.9-2.15).$ 
  Further interactions generate a  low density  upper zone, also formed by the coupling of nonlinear upward and downward moving perturbations, it is separated from the later zone by a sharp curve. The variation of the density   as a function of the sunward direction for fixed times  can also be seen from Fig.~\ref{fig:dos}c. For the non--dimensional time  $t_{n}=0.594$ we note a first peak  corresponding  to the main shock front, the posterior smooth decaying values of density corresponds to the expansion region. The interaction with descending perturbations determine another shock peak with a descending step of  density at $\sim 0.41.$ The behaviour is similar for  $t_{n}=0.95$ without the second shock. At larger times, $t_{n}=1.38,$ as the main shock was already absorbed, only the  step of the density, while crossing the voided region, is registered.

\begin{figure}
 \includegraphics[width=8.3cm]{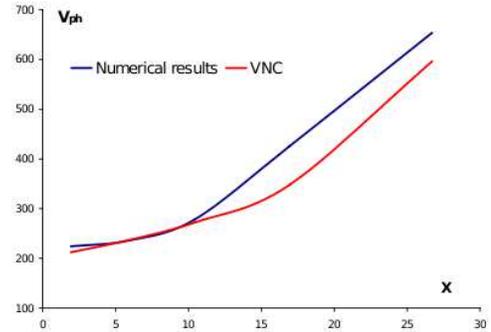}

\caption{Numerical expansion curve vs. observational VNC data. At $x=x_{o}$, origin of coordinates, $V_{ph}=V_{pho}$}
  \label{fig:tres}
   \end{figure}

To compare our speed values with those in VNC we calculated the speed of the  curve  limiting the vacuum zone. The VNC curve was obtained with the speed function of eq.~(\ref{Amp}) replacing  $x_{o}$ by $0$, in accordance with our choice of the origin of coordinates. Figure~\ref{fig:tres} shows the comparison between the numerical and observational   curves.  Note that the the reference value $V_{pho}=202$kms$^{-1}$, the speed measured by VNC at $x_{o},$ is the simulation speed  at the origin of coordinates.
The figure shows the good correspondence between the observational velocity curve and the numerical one.

\begin{table}
\begin{tabular}{ccccccc}
&$Edge$& $\tau  (s)$ &$ V_{pho} (km  s^{-1})  $ &$ L_{v} (Mm)  $ & $  A_{o}  (km)$&$ L_{A} (Mm)  $
\\  \hline \hline
&$C$&$134$&$202$&$40.4$&$906$&$34$
\\ \hline
&$D$&$175$&$ 176$&$14.3$&$246$&$11.5$
\\ \hline
 \hline
\end{tabular}
\caption{\label{tab:table1} Observational wave parameters  taken from VNC.  }
\end{table}



\begin{table}
\begin{tabular}{ccccc}
&$Edge$&$y_{1},y_{2},y_{3}$&$P_{2}/P_{1,3} $&$x_{1},x_{2},x_{3}$
\\  \hline \hline
&$C$&$0.7,0.04,0.26$&$10$& --
\\ \hline
&$D$&$0.48,0.04,0.48$&$10$& --
\\ \hline
&$C (radial \ case)$&--&$10$&$0.7,0.25,0.05$
\\ \hline
\end{tabular}
\caption{\label{tab:table2} Initial condition values for the transverse simulation shown in Fig.~\ref{fig:uno}; first two lines. Initial condition values for the sunward simulation shown in Fig.~\ref{fig:dos}; third line.  $P_{1}=P_{3}, \ \rho_{1}=\rho_{2}=\rho_{3}=1.$}
\end{table}

\section{Summary }

We have integrated the MHD ideal equations   to simulate observational dark lane data obtained by Verwichte et al. \cite{ver} (VNC). 
We simulated the effects of an initial impulsive and localized deposition of  energy -supposed to be associated with above reconnection processes-  in a plasma structured by sunward magnetic field lines.    The impulsive phase is modeled by a pressure perturbation that initiates two main different type of processes, a fundamentally hydrodynamic shock pattern directed sunwards and a perpendicular magnetic shock one, i.e., transversal to the magnetic field. The two patterns are supposed to be  independent processes however linked by their common origin and background magnetic and density conditions. 
Firstly, we reproduced the oscillatory kink--like pattern of the transverse $y$ component. Varying the initial conditions we were able to calibrate and  reproduce  the  C edge  in VNC. Also a sausage--like pattern was obtained resembling the tail edge D  in VNC. The differences between patterns are due to the  non--symmetric (most probable kink--like) or   symmetric (less probable sausage--like) rebounds of the magnetic transverse shocks in the denser external medium. The resulting interactions of  bounced nonlinear shocks  composes and sustains, in accordance with observational characteristic times of the phenomenon, a density structure with  a central  void resembling a  kink--like mode or a sausage--like mode as shown in Fig.~\ref{fig:uno}a-b. 

The same reference values were used  as  initial condition to simulate the sunward evolution. We could qualitatively reproduce the observational data showing that initially two opposite  shock  evolving fronts are produced. One    evolves towards the sun surface until is absorbed, and the other  is forced to rebound upwardly resembling  the action of a reconnection site. Afterwards, the interaction of upwardly and downwardly moving perturbations form an expansion wave region that lowers the density of the medium. Also a voided zone is formed and 
 sustained,  lasting for times comparable with observational times, due to the continued interfering of non--linear waves. This evolution is described in  Fig.~\ref{fig:dos} which reproduces with good accuracy   Fig.~(2C) 
 in  VNC. 
Finally, the composition of   the two   dynamics,  one corresponding to the sunward directed hydrodynamic shock pattern  and the  perpendicular magnetic shock, composes an overall description of  a moving  transversally shaking void, resembling  a kink--like mode as it moves towards the sun surface. 

\begin{acknowledgements}
We are grateful to the unknown referee for his/her comments and
suggestions that helped to understand and improve the quality of the paper.
\end{acknowledgements}


\end{document}